\documentclass[a4paper,11pt]{article}
\usepackage{pos}

\usepackage{pos}
\usepackage[english]{babel} %per ridefinire convenzioni italiane
\usepackage[utf8]{inputenc}
\usepackage{amsthm} %per classe theorem etc
\usepackage{mathtools}
\usepackage{pgfplots}
\usepackage{tensor} %per indici tensori
\usepackage{physics}
\usepackage{bm}
\usepackage{feynmp}
\usepackage{feynmp-auto}

\newcommand{\reffig}[1]{Fig.~\ref{#1}}
\newcommand{\refsec}[1]{Sec.~\ref{#1}}

\newcommand{\eqrefeq}[1]{Eq.~\eqref{#1}}

\title{Chebyshev and Backus-Gilbert reconstruction for inclusive semileptonic $B_{(s)}$-meson decays from Lattice QCD}
\ShortTitle{Chebyshev and Backus-Gilbert reconstruction for inclusive semileptonic decays}

\author*[a,b,c]{Alessandro Barone}
\author[e,f]{Shoji Hashimoto}
\author[b,c,d]{Andreas J\"uttner}
\author[e,f,g]{Takashi Kaneko}
\author[e,f]{Ryan Kellermann}

\affiliation[a]{PRISMA+ Cluster of Excellence \& Institut f\"ur Kernphysik, Johannes-Gutenberg-Universit\"at Mainz, D-55099 Mainz, Germany}
\affiliation[b]{School of Physics and Astronomy, University of Southampton, Southampton SO17 1BJ, UK}
\affiliation[c]{STAG Research Center, University of Southampton, Southampton SO17 1BJ, UK}
\affiliation[d]{CERN, Theoretical Physics Department, Geneva, Switzerland}
\affiliation[e]{High Energy Accelerator Research Organization (KEK), Ibaraki 305-0801, Japan}
\affiliation[f]{School of High Energy Accelerator Science, SOKENDAI (The Graduate University for Advanced Studies), Ibaraki 305-0801, Japan}
\affiliation[g]{Kobayashi-Maskawa Institute for the Origin of Particles and the Universe, Nagoya University, Aichi 464–8602, Japan}

\emailAdd{abarone@uni-mainz.de}

\abstract{
We present a study on the nonperturbative calculation of observables for inclusive semileptonic decays of $B_{(s)}$ mesons using lattice QCD.
We focus on the comparison of two different methods to analyse the lattice data of Euclidean correlation functions, specifically Chebyshev and Backus-Gilbert approaches. This type of computation may eventually provide new insight into the long-standing tension between the inclusive and exclusive determinations of the Cabibbo-Kobayashi-Maskawa (CKM) matrix elements $|V_{cb}|$ and $|V_{ub}|$.
We report the results from a pilot lattice computation for the decay $B_s \rightarrow X_c \, l\nu_l$, where the valence quark masses are approximately tuned to their physical values using the relativistic-heavy quark action for the $b$ quark and the domain-wall formalism for the other valence quarks. We address the computation of the total decay rate as well as leptonic and hadronic moments, discussing similarities and differences between the two analysis techniques.
}

\FullConference{The 40th International Symposium on Lattice Field Theory (Lattice 2023)\\
July 31st - August 4th, 2023\\
Fermi National Accelerator Laboratory\\}

\begin{document}
\maketitle

\section{Introduction}

Quark-flavour physics is an area of particular interest to search for deviation from the Standard Model (SM), as
weak processes characterised by flavour-changing currents  may be sensitive to New Physics: 
while new particles may be too heavy to be produced with energies achievable by current experimental facilities,
quantum effects could leave detectable traces in flavour-physics processes. Therefore, any discrepancy
between SM theoretical predictions and experimental measurements could be an indicator of new effects.

One intriguing puzzle is the long-standing tension between the inclusive and exclusive
determinations of the Cabibbo-Kobayashi-Maskawa (CKM) matrix elements $|V_{cb}|$ and
$|V_{ub}|$. Semileptonic decays of $B$ mesons constitute the main channel for the
extraction of these parameters, and therefore represent crucial processes to address and
investigate this tension.

First viable theoretical
proposals for how to accomplish the computation of inclusive decay observables on the
lattice have appeared only recently~\cite{Hansen2017,Hashimoto2017,Gambino2020}.
We report on our pilot study on the calculation of the decay rate of the
inclusive semileptonic $B_s \rightarrow X_c \, l\nu_l$ decay~\cite{Barone:2022gkn,Barone:2023tbl}, focusing on similarities and differences of the analysis
strategy based on the Chebyshev polynomial and Backus-Gilbert reconstruction.
We present further work on the generalisation of our setup to the computation of moments of various kinematical quantities, in particular hadronic mass
and lepton energy moments.

\section{Inclusive decay rate and kinematic moments}

The starting point of the calculation is given by the differential decay rate
\begin{align}
\label{eq:diff_decay_rate}
 \frac{\dd \Gamma}{\dd q^2 \dd q_0 \dd E_{l}} = \frac{G^2_F |V_{cb}|^2}{8\pi^3} L_{\mu\nu} W^{\mu\nu} \, ,
\end{align}
where $W^{\mu\nu}\equiv W^{\mu\nu}(p, q) $ is the hadronic tensor for the $B_s$ decay defined as
\begin{align}
 W^{\mu\nu} = \sum_{X_c}(2\pi)^3 \delta^{(4)}(p-q-r)\frac{1}{2E_{B_s}(\bm{p})} 
 \bra{B_s(\bm{p})} J^{\mu\dagger}(0)\ket{X_{c}(\bm{r})}  \bra{X_{c}(\bm{r})} J^{\nu}(0) \ket{B_s(\bm{p})} \, ,
 \label{eq:hadronicTensor}
\end{align}
which contains all the nonperturbative QCD effects,
and $L^{\mu\nu}$ is the leptonic tensor
%
%\begin{align}
% L^{\mu\nu} = p_{l}^{\mu}p_{\nu_l}^{\nu} +  p_{l}^{\nu}p_{\nu_l}^{\mu} - g^{\mu\nu} p_{l}\cdot p_{\nu_l} -i\epsilon^{\mu \alpha \nu \beta} p_{l\,,\alpha }p_{\nu_l\,,\beta} \, .
%\end{align}
%%
which contains known kinematic terms associated with the lepton-neutrino pair.

The total decay rate is obtained integrating~\eqrefeq{eq:diff_decay_rate}
\begin{align}
 \Gamma &= \frac{G_F^2 |V_{cb}|^2}{24\pi^3} \int_{0}^{\bm{q}^2_{\rm max}} \dd \bm{q}^2\, \sqrt{\bm{q}^2} \, \bar{X}(\bm{q}^2)\, , \qquad
 \label{eq:total_Xbar}
\end{align}
where we changed the integration variables to the three-momentum $\bm{q}^2$ of the hadronic state $X_c$ and its energy $\omega$, with
\begin{align}
 \bar{X}(\bm{q}^2) \equiv    \frac{3}{\sqrt{\bm{q}^2}} \int_{\omega_{\rm 0}}^{\infty}\dd \omega \int_{E_l^{\rm min}}^{E_l^{\rm max}} \dd E_l  L_{\mu\nu} W^{\mu\nu}
 = \int_{\omega_{\rm 0}}^{\infty} \dd \omega \, K_{\mu\nu} W^{\mu\nu} \, ,
 \label{eq:kernel_decay}
\end{align}
where $K_{\mu\nu}=K_{\mu\nu}(\bm{q}^2, \omega)$ is a kernel function that contains known kinematic factors obtained from the leptonic tensor after
the integration over the lepton energy $E_l$.
While the energy phase space in $\omega$ is restricted in some finite interval $\omega \in [\omega_{\rm min}, \omega_{\rm max}]$ for every $\bm{q}^2$, 
we are free to modify the integration range to $\omega_0 \leq \omega_{\rm min}$, since the hadronic tensor has no support below the ground state $\omega_{\rm \min}$, 
and extend $\omega_{\rm max}\rightarrow \infty$ including a Heaviside function into the kernel $K_{\mu\nu}$. These modifications
will be relevant later for the analysis strategy.

Moving to the moments of a given kinematic quantity $p$ at order $n$, these are defined as
\begin{align}
 \langle (p)^n \rangle &= \frac{\Gamma_{p_n}}{\Gamma}  \, , \quad
 \Gamma_{p_n}  \equiv  \int \dd \bm{q}^2 \dd \omega \dd E_l \, (p)^n \left[ \frac{\dd\Gamma}{\dd\bm{q}^2\,\dd\omega\,\dd E_l} \right] \, ,
\end{align}
which can be rewritten as
\begin{align}
\label{eq:moments}
 \langle (p)^n \rangle &=
 \frac{\int \dd \bm{q}^2 \sqrt{\bm{q}^2} \bar{X}^{(n)}_p(\bm{q}^2) }
 { \int \dd \bm{q}^2 \sqrt{\bm{q}^2} \bar{X}(\bm{q}^2) } \, ,
\end{align}
with 
\begin{align}
 \bar{X}^{(n)}_p(\bm{q}^2)= \frac{3}{\sqrt{\bm{q}^2}} \int_{\omega_{\rm 0}}^{\infty}\dd \omega \int_{E_l^{\rm min}}^{E_l^{\rm max}} \dd E_l \, (p)^{n} L_{\mu\nu} W^{\mu\nu}
  =\int_{\omega_{\rm 0}}^{\infty} \dd \omega \, K^{(n)}_{p,{\mu\nu}} W^{\mu\nu} \, ,
  \label{eq:kernel_moments}
\end{align}
where $K^{(n)}_{p,\mu\nu}$ is the corresponding kernel function analogous to the one in~\eqrefeq{eq:kernel_decay}.
We consider in particular the hadronic mass $(H)$ moments  $\langle (M_X^2)^n \rangle$, with $M_X^2=(\omega^2-\bm{q}^2)$,
and the lepton energy $(L)$ moments $\langle (E_l)^n \rangle$.
In order to discuss and illustrate our method,
we will focus on the computation of $\bar{X}(\bm{q}^2)$ and the analogous quantities for the $n=1$ moments $\bar{X}^{(1)}_H(\bm{q}^2)$ and $\bar{X}^{(1)}_L(\bm{q}^2)$,
which are the crucial ingredients to compute the final observables of interest.

\section{Lattice approach}

Inclusive decays can be studied on the lattice through the computation of four-point correlation functions
\begin{align}
\label{eq:4pt}
 C_{\mu \nu}^{S J J S}\left(t_{\rm snk}, t_{2}, t_{1}, t_{\rm src}\right) =&
 \sum_{\bm{x}_{\rm snk}} e^{-i \bm{p}_{\rm snk}\cdot (\bm{x}_{\rm snk}-\bm{x}_{\rm src})} 
 \left\langle T 
 \left\{ \mathcal{O}_{B_s}^{S}\left(x_{\rm snk}\right) 
 \tilde{J}_{\mu}^{\dagger}\left(\bm{q}, t_{2}\right) \tilde{J}_{\nu}\left(\bm{q}, t_{1}\right) 
   \mathcal{O}_{B_s}^{S \dagger}\left(x_{\rm src} \right) \right\} \right\rangle  \, ,
\end{align}
where $\mathcal{O}_{B_s}$ is an interpolating operator for the the $B_s$ meson,
and $\tilde{J}_{\nu}$ is the $\bar{b}\rightarrow \bar{c}$ weak current projected in momentum space.
In particular, the hadronic tensor can be addressed through the ratio
\begin{equation}
 C_{\mu\nu}(\bm{q}, t) \equiv
 \frac{1}{2M_{B_s}}|\bra{0} \mathcal{O}_{B_s}^{L} \ket{B_s} |^2 
 \frac{C^{SJJS}(t_{\rm snk},t_2 , t_1, t_{\rm src})}{C^{SL}(t_{\rm snk}, t_2)C^{LS}(t_1, t_{\rm src})} \,,
\label{eq:Cmunu}
\end{equation}
with $t_{\rm snk}-t_2 \gg 1$, $t_1-t_{\rm src}\gg 1$ and $t=t_2-t_1$, and where $C^{XY}$ represent the two-point $B_s$ correlator, with
the superscripts indicating smearing $(S)$ or no smearing $(L)$. Indeed, the new correlator $C_{\mu\nu}(\bm{q}, t)$
is related to the hadronic tensor as
\begin{align}
 \begin{split}
 C_{\mu\nu} (\bm{q}, t)
% &= \int_{\omega_0}^{\infty} \dd \omega \, \frac{1}{2M_{B_s}}
%    \bra{B_s} \tilde{J}_{\mu}^{\dagger}(\bm{q},0) \delta (\hat{H} - \omega) \tilde{J}_\nu (\bm{q},0) \ket{B_s} e^{-\omega t} \\
 &=  \int_{\omega_0}^{\infty} \dd \omega \, W_{\mu\nu}(\bm{q}, \omega) e^{-\omega t} \, ,
 \end{split}
\end{align}
which shows that $W_{\mu\nu}$ represents the spectral function for the correlator $C_{\mu\nu}$ in the K\"all\'en-Lehmann
representation, namely
\begin{align}
 W_{\mu\nu}(\bm{q}, \omega) = \frac{1}{2M_{B_s}}\sum_{X_c} \delta(\omega-E_{X_c}) \bra{B_s} \tilde{J}_{\mu}^{\dagger}(\bm{q},0)\ket{X_c}\bra{X_c} \tilde{J}_{\nu}(\bm{q},0)\ket{B_s} 
\end{align}
in the rest frame of the $B_s$ meson.

The extraction of the hadronic tensor requires the computation of the inverse Laplace transform and represents therefore an ill-posed inverse problem.
However, for the calculation of the quantities $\bar{X}(\bm{q}^2)$ for both decay rate and moments the extraction of the spectral function can be
bypassed and the observables can be evaluated directly. In particular, referring to Eqs.~\eqref{eq:kernel_decay} and~\eqref{eq:kernel_moments},
$\bar{X}(\bm{q}^2)$ can be obtained naively through a polynomial approximation in $e^{-\omega}$ (in lattice units) of the kernel function up to a degree $N$,
i.e. $K_{\mu\nu} = \sum_{j=0}^{N} c_{\mu\nu,j} e^{-j\omega} $ such that
\begin{align}
 \bar{X}_{\rm naive}(\bm{q}^2) = \sum_{j=0}^{N} c_{\mu\nu,j} \int_{\omega_0}^{\infty} \dd \omega \, W^{\mu\nu} e^{-j\omega} 
                  = \sum_{j=0}^{N} c_{\mu\nu,j} C^{\mu\nu}(j) \, .
\end{align}
Note that the approximation of the kernel is possible as long as the Heaviside function contained in $K_{\mu\nu}$ is regularised with a continuous function
such as a sigmoid $\theta_{\sigma}(x)=1/(1+e^{-x/\sigma})$,
where $\sigma$ is a parameter that controls the sharpness of the step and approaches the Heaviside for $\sigma \rightarrow 0$. While this
observation plays an important role in the final calculation, we won't discuss it further and refer to other works for more details~\cite{Gambino2022,Barone:2023tbl}.
We labelled the above calculation as ``naive'' because, while theoretically well-defined, such a procedure would in practice lead to a build up
of the statistical noise originating from the correlator at different time slices, such that the final error on $\bar{X}_{\rm naive}(\bm{q}^2)$
would be too large to make any significant phenomenological prediction. To proceed in the analysis we then need to consider two steps:
\begin{enumerate}
 \item find a suitable polynomial approximation strategy to approximate the kernels;
 \item devise a ``regularisation'' to reduce the variance of the target observable.
\end{enumerate}
The final result for the observable is then
\begin{align}
 \bar{X}(\bm{q}^2) = \bar{X}_{\rm naive}(\bm{q}^2) + \delta \bar{X}(\bm{q}^2) \, ,
\end{align}
where $\delta\bar{X}(\bm{q}^2)$ is the regularisation term which acts as a noisy zero that does not change the final result
but takes care of reducing the variance.

\section{Chebyshev and Backus-Gilbert reconstruction}
\label{sec:recon}

\begin{figure}[t]
\centering
\hbox{
\includegraphics[scale=0.3]{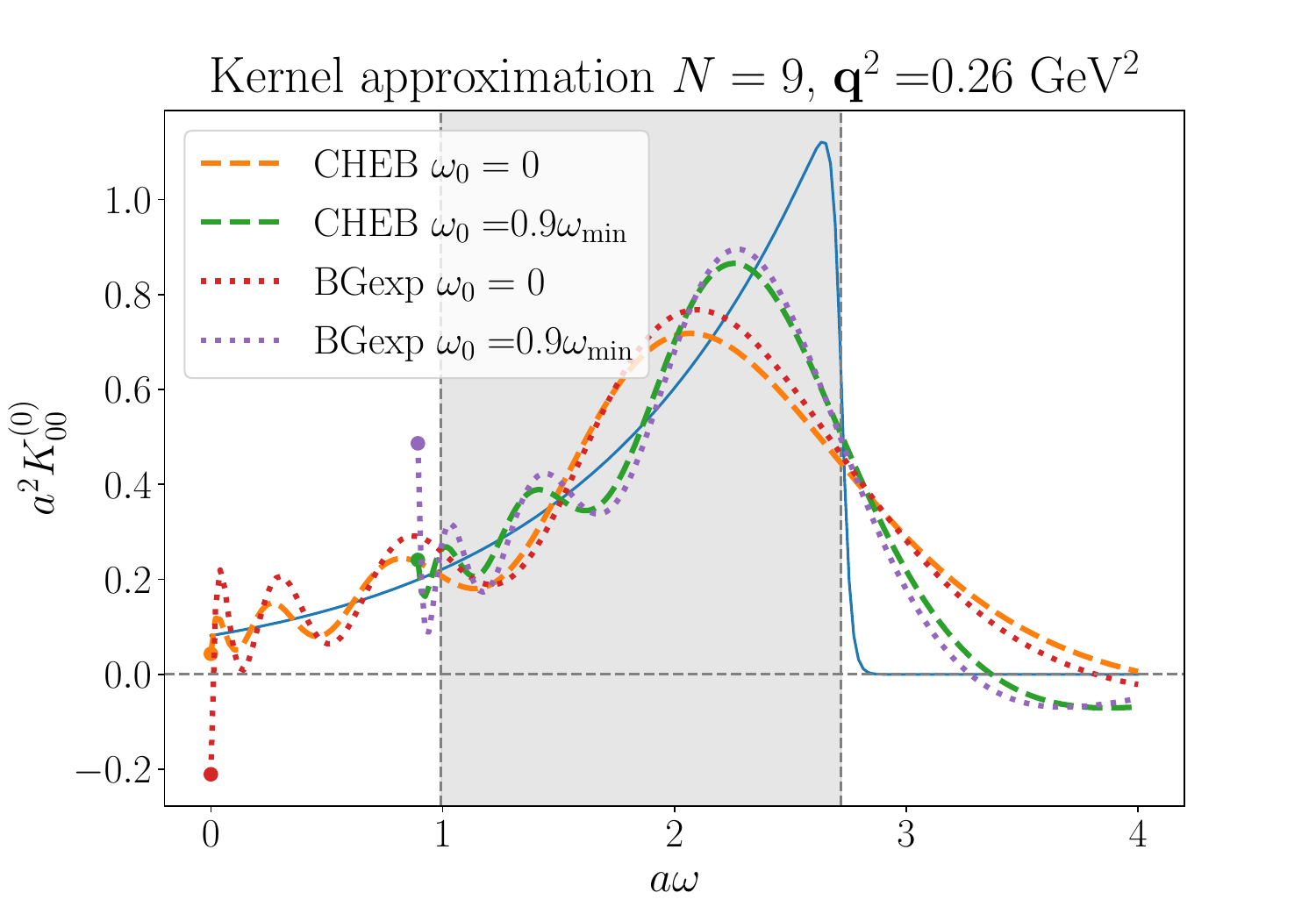}
\hspace{-0.5cm}
\includegraphics[scale=0.3]{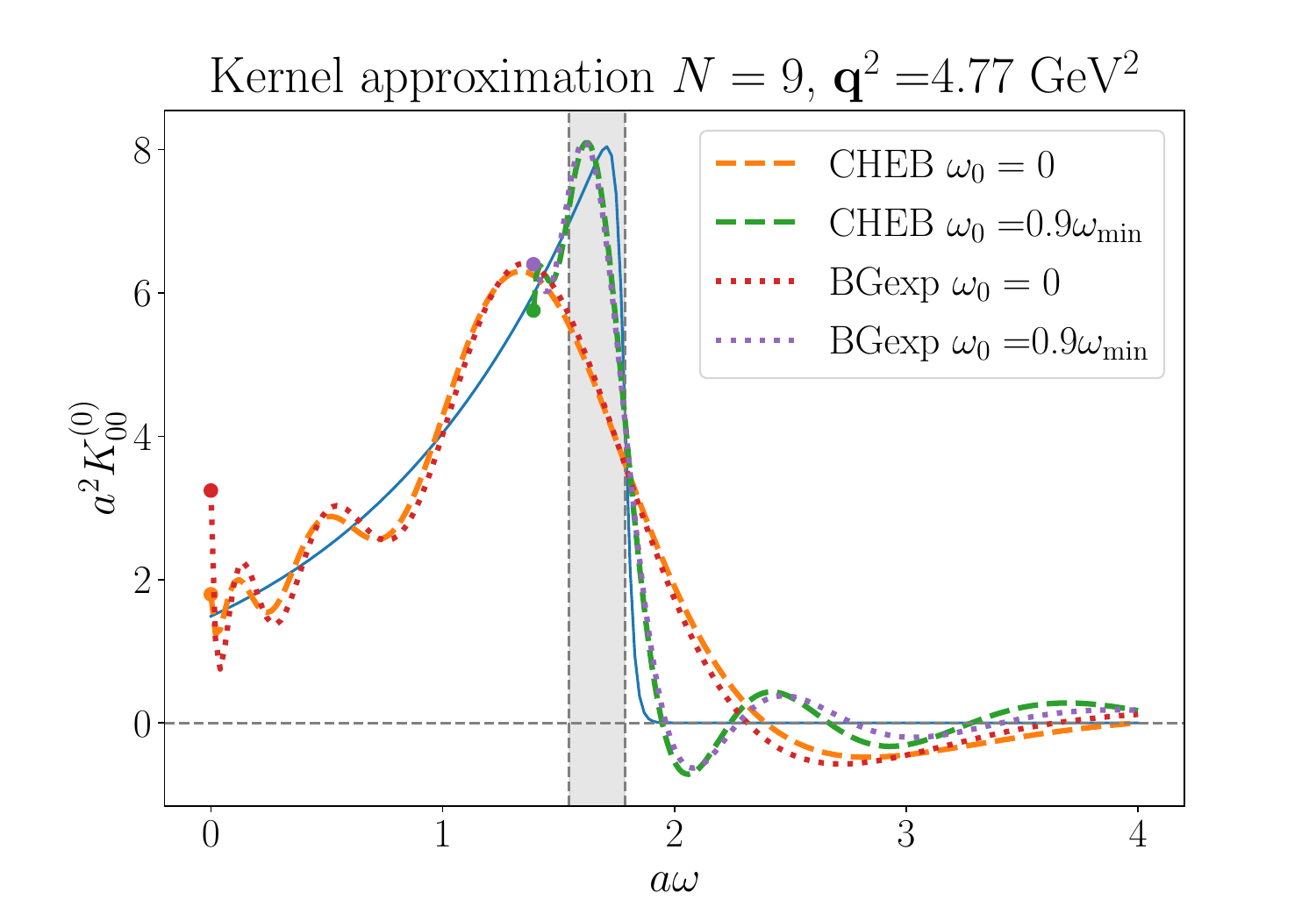}
}
\caption{Example of the polynomial approximation of one of the kernels for the decay rate at degree $N=9$ with Chebyshev and Backus-Gilbert method
with two different starting point $\omega_0$ for the polynomial approximation.}
\label{fig:pol_app}
\end{figure}

For the analysis strategy we rely on polynomial approximations based on the Chebyshev polynomial technique~\cite{Barata:1990rn,Weisse2006}
and the Backus-Gilbert method~\cite{Backus1968,Hansen2019}. We refer to the App. of~\cite{Barone:2023tbl} for a deeper discussion of both approaches.
Here we point out the main differences between the two in terms of the polynomial approximation of the kernels (with no connection to the data)
as well as the variance reduction of the final $\bar{X}(\bm{q}^2)$ observables.

Let us start from the polynomial approximation. 
We first highlight that, by visual inspection of~\reffig{fig:pol_app}, the quality of the two polynomial
approximation strategies is comparable. Furthermore, the choice of the starting point of the approximation $\omega_0$ (cf. with the integral in~\eqref{eq:kernel_decay}) has a significant impact on the quality of the reconstruction, in particular for larger $\bm{q}^2$, where the phase
space in $\omega$ is shrunk (right plot).

Concerning the variance reduction, the two methods act in a significantly different way. The Chebyshev-polynomial technique relies
on trading the original data with a refitted set that accounts for rigorous mathematical bounds associated with the shifted Chebyshev polynomials 
$\tilde{T}_{j}(\omega)$, $\omega \in [\omega_0, \infty)$, for which $|\tilde{T}_{j}(\omega)|\leq 1$. Namely, the normalised correlator in~\eqrefeq{eq:Cmunu}
can be written as linear combination of
$\tilde{T}_{j, \mu\nu} \equiv \left(\int_{\omega_0}^{\infty}\dd \omega W_{\mu\nu} \tilde{T}_{j}(\omega)e^{-2t_0\omega}\right) /
\left( \int_{\omega_0}^{\infty}\dd \omega W_{\mu\nu} \tilde{T}_{0}(\omega)e^{-2t_0\omega} \right) $,
such that we can obtain a new set of data for the normalised correlator
$\bar{C}_{\mu\nu}(\bm{q}, j) =  C_{\mu\nu}(\bm{q}, j+2t_0) / C_{\mu\nu}(\bm{q}, 2t_0)$ requiring
\begin{align}
 \bar{C}^{\rm fit}_{\mu\nu}(\bm{q}, k)  = \sum_{j=0}^{k} \tilde{a}^{(k)}_{j} \tilde{T}_{j, \mu\nu} \, ,
\end{align}
and imposing the bounds $|\tilde{T}_{j,\mu\nu}|\leq 1$ in the fitting procedure, where $\tilde{a}^{(k)}_{j}$ is a set of known coefficients associated with the power
representation of the polynomials. The shift $t_0$ represents
the minimum distance between the two currents in~\eqrefeq{eq:4pt} and has to be accounted for in $K_{\mu\nu}$. In this way, the correction term reads
\begin{align}
 \delta\bar{X}(\bm{q}^2) = C_{\mu\nu}(\bm{q}, 2t_0) \sum_{j=0}^{N} c_{\mu\nu, j} \, \delta \bar{C}_{\mu\nu}(\bm{q}, j) \, , \quad
 \delta \bar{C}_{\mu\nu}(\bm{q}, j) = \bar{C}^{\rm fit}_{\mu\nu}(\bm{q}, j) - \bar{C}_{\mu\nu}(\bm{q}, j)
\end{align}
and the variance is minimised acting of the correlator data. 

On the other hand, the Backus-Gilbert method reduces the variance by modifying the coefficients of the polynomial approximation such that
they account also for the reduction of the statistical noise. This is achieved in practice by minimising the functional
\begin{align}
 F = A + \theta^2 B \, ,
\end{align}
where the functional $A$ addresses the pure polynomial approximation (systematic error) with a chosen polynomial basis $b_j(\omega)$,
$B$ addresses the variance of $\bar{X}(\bm{q}^2)$ (statistical error) and $\theta^2$
is a parameter chosen by hand that accounts for the interplay between the two types of error. The ``naive'' result is then regularised
through a correction to the coefficients as
\begin{align}
 \delta\bar{X}(\bm{q}^2) = C_{\mu\nu}(\bm{q}, 2t_0)  \sum_{j=0}^{N} \delta c_{\mu\nu, j} \, \bar{C}_{\mu\nu}(\bm{q}, j) \, , \quad
 \delta c_{\mu\nu, j} = \delta c_{\mu\nu, j}\Big|_{\theta^2\neq 0} - \delta c_{\mu\nu, j}\Big|_{\theta^2=0} \, .
\end{align}

\section{Numerical setup}

Our calculation is based on a $24^3\times 64$ lattice with 2+1-flavour domain-wall fermion (DWF)~\cite{Shamir1993,Furman1994}
gauge-field ensembles with the Iwasaki gauge action~\cite{Iwasaki:1983iya} from the RBC/UKQCD Collaboration \cite{Allton2008}
at lattice spacing $a^{-1}=1.785(5) \, \text{GeV}$ (corresponding to $a\simeq 0.11 \,\text{fm}$), pion mass $M_{\pi}\simeq 340 \,\text{MeV}$
and close-to-physical strange-quark mass.
The computations have been performed with the Grid \cite{Grid,GridProc,Yamaguchi:2022feu} and Hadrons \cite{HadronsZenodo} software packages.

We use the same simulation parameters RBC/UKQCD is using in the heavy-light meson projects on exclusive semileptonic $B_{(s)}$ meson decays~\cite{Flynn2018,Flynn2019,Flynn2021,Flynn:2023ufa}.
In particular, the valence-strange quark is simulated using DWF, whereas the valence-charm quark is simulated by using the M\"obius DWF action \cite{Cho2015,Brower2017}.
Their masses are tuned such that mesons containing bottom, charm and strange valence quarks have masses close to the physical ones.
The physical bottom quark cannot currently be simulated with DWF and some EFT-based action is required: the
$b$ quark has been simulated at its physical mass using the Columbia formulation of the relativistic-heavy-quark (RHQ) action \cite{RHQColumbia1,RHQColumbia2},
which is based on the Fermilab heavy quark action \cite{RHQFermilab}. 

For the computation we average over 120 statistically independent gauge configurations, and on each configuration the measurements are performed on 8 different linearly spaced source-time planes.
We use $\mathbb{Z}_2$ wall sources~\cite{Foster:1998vw,McNeile:2006bz,Boyle:2008rh} to improve the signal.
We compute 8 different momenta linearly spaced in $\bm{q}^2$ to cover the full kinematic range and another 2 momenta to increase the resolution
for small $\bm{q}^2$. These have been induced through partially twisted boundary conditions \cite{DeDivitiis,Sachrajda2004}
for the charm quark with the same momentum in all three spatial directions. 
 
\section{Results}

\begin{figure}[t]
\centering
\hbox{
\includegraphics[scale=0.3]{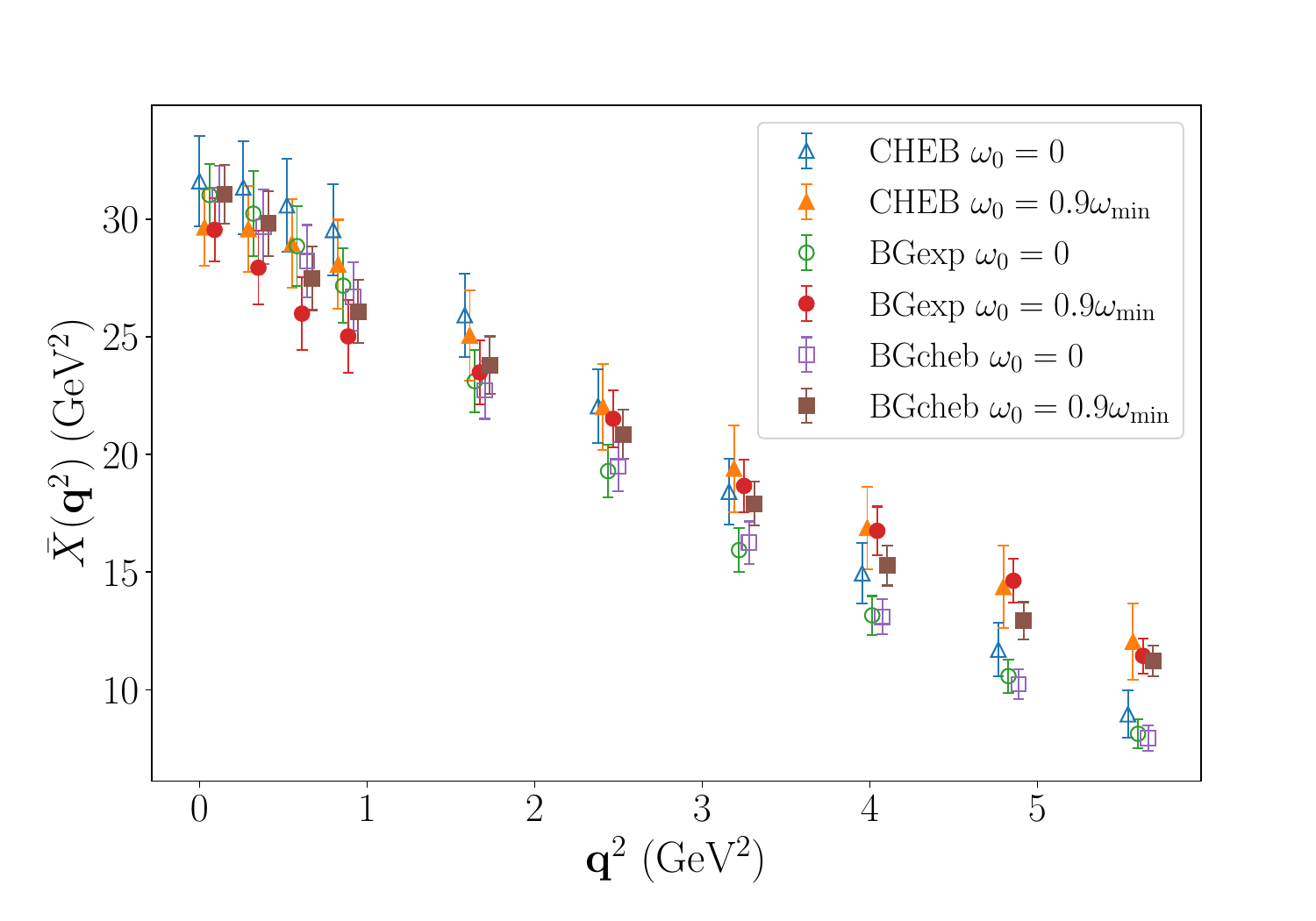}
\hspace{-0.5cm}
\includegraphics[scale=0.3]{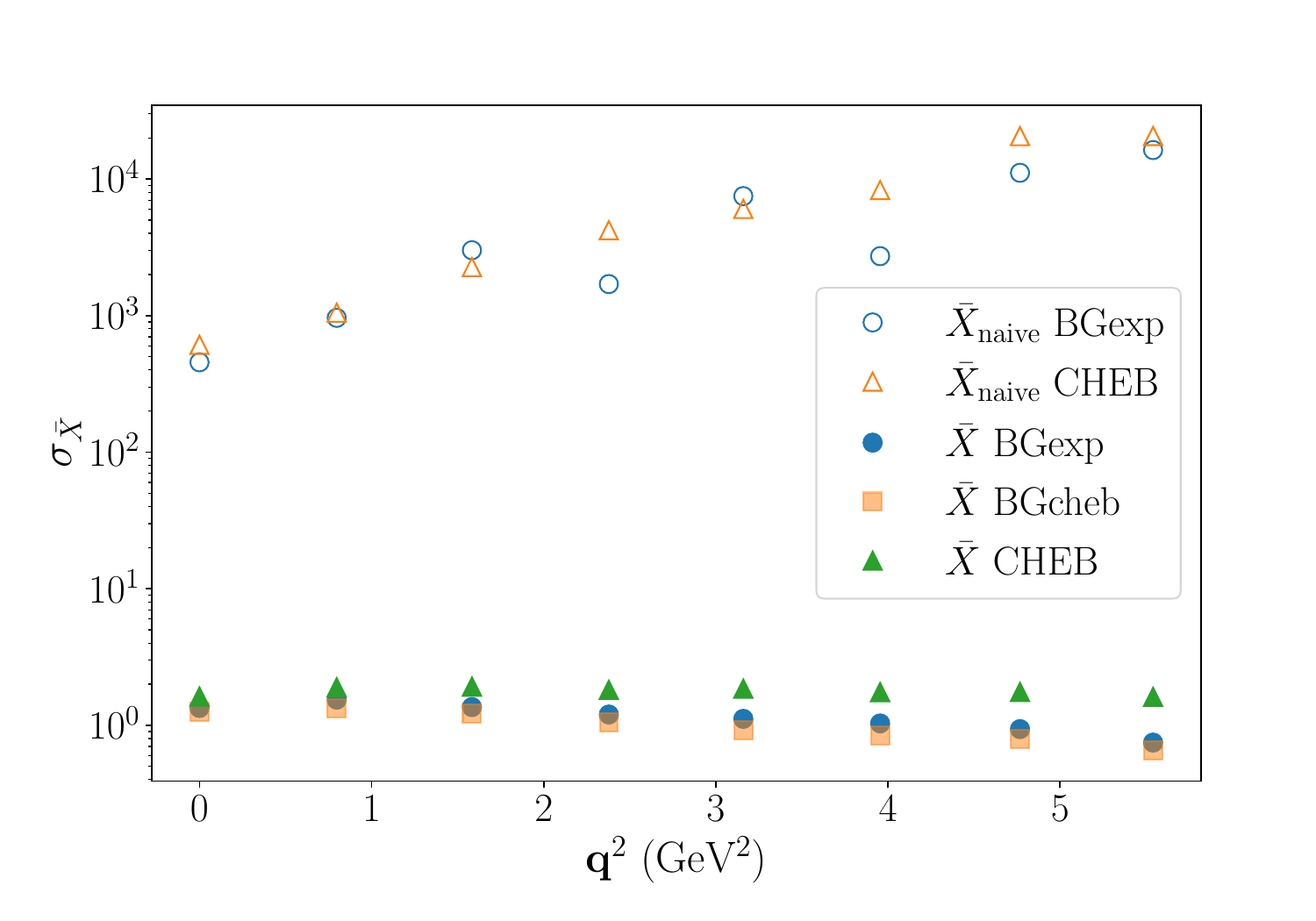}
}
\caption{Left: estimate of $\bar{X}(\bm{q}^2)$ with the two different strategies for 10 different values of $\bm{q}^2$
with $N = 9$ and $\bm{q}_{\rm max}= 5.83 \,\text{GeV}^2$. 
Right: effect of the variance reduction to $\bar{X}_{\rm naive}(\bm{q}^2)$ from the correction $\delta\bar{X}(\bm{q}^2)$
for the case $\omega_0=0.9\omega_{\rm min}$.
The y axis shows the standard deviation $\sigma_{\bar{X}}$ for $\bar{X}_{\rm naive}(\bm{q}^2)$ (empty symbols) and
$\bar{X}(\bm{q}^2)=\bar{X}_{\rm naive}(\bm{q}^2)+\delta\bar{X}(\bm{q}^2)$ (filled symbols).}
\label{fig:Xbar}
\end{figure}

We now present some of the main findings of this study. For all the observables we report results from the two reconstruction approaches
at values $\omega_0=0$ and $\omega_0=0.9 \omega_{\rm min}$. For the Backus-Gilbert method, we employ two different
choices for the polynomial basis, namely exponentials $b_{j}(\omega)=e^{-j\omega}$ or shifted Chebyshev polynomials $b_{j}(\omega)=\tilde{T}_{j}(\omega)$.

In~\reffig{fig:Xbar} we show the results for the
quantity $\bar{X}(\bm{q}^2)$ (left) together with the effect of the variance reduction (right) outlined in~\refsec{sec:recon}.
In particular, the latter shows that the reduction of the statistical error is substantial and it therefore emphasises the importance
of regularisation methods for the final result.
All the methods are in agreement, and the only deviation is associated with the different values of $\omega_0$ as $\bm{q}^2$ increases.
This is understood in terms of the different quality of the approximation as the phase space in $\omega$ shrinks, as illustrated in~\reffig{fig:pol_app},
and has to be taken into account in the systematics.

\begin{figure}[t]
\centering
\hbox{
\includegraphics[scale=0.3]{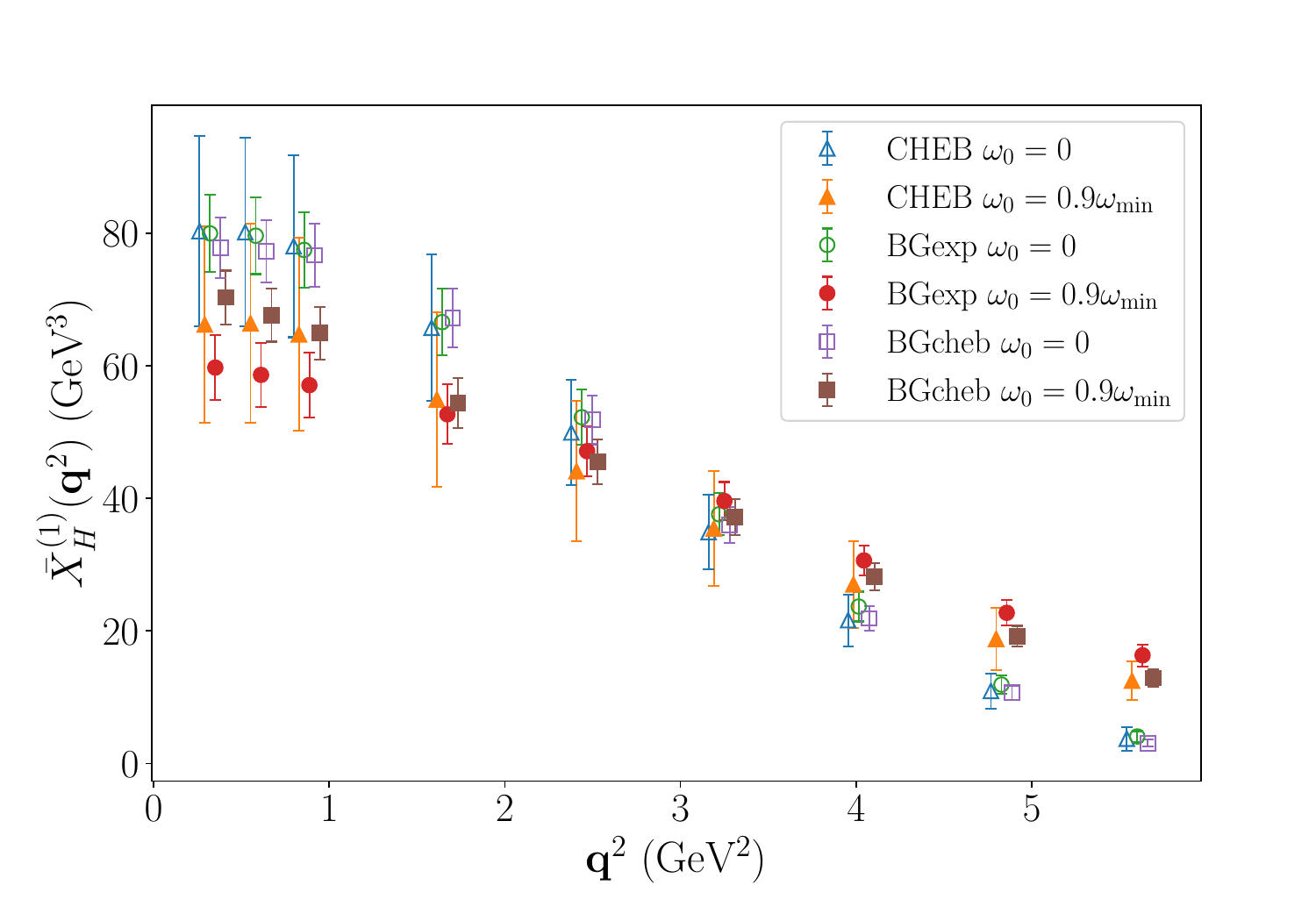}
\hspace{-0.5cm}
\includegraphics[scale=0.3]{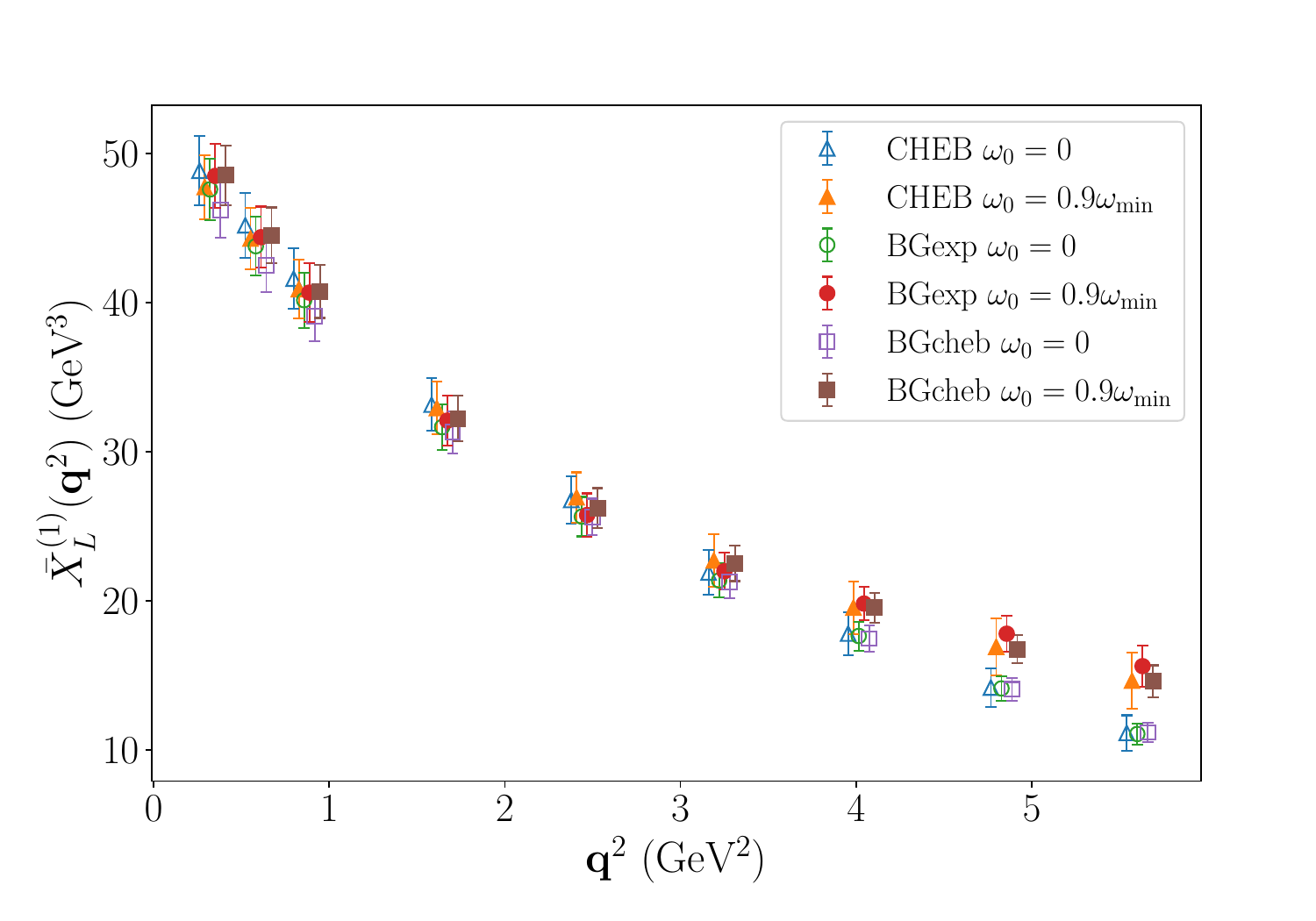}
}
\caption{Evaluation of the numerators $\bar{X}^{(1)}_{H}(\bm{q}^2)$ (left) and $\bar{X}^{(1)}_{L}(\bm{q}^2)$ (right) of
the hadronic mass and lepton differential moments at $n=1$, respectively.}
\label{fig:Xbar_mom}
\end{figure}

In~\reffig{fig:Xbar_mom} we show some of the results for the moments. In particular, we focus on
the numerators $\bar{X}^{(n)}_{H}(\bm{q}^2)$ and $\bar{X}^{(n)}_{L}(\bm{q}^2)$  that enter
the definition of the hadronic mass and lepton moments at order $n=1$, as in the general expression~\eqrefeq{eq:moments}.
As for the decay rate, we see excellent agreement among the different methods.
Note that, compared to the decay rate, the errors are
larger for $\bar{X}^{(1)}_{H}(\bm{q}^2)$ and smaller (or comparable) for $\bar{X}^{(1)}_{L}(\bm{q}^2)$. This can be understood
in terms of the differences of the kernels: indeed, the hadronic mass moments introduce
extra factors that depend on $\omega$ in the kernels, whereas the behaviour of the leptonic
kernels is smooth. We therefore expect the polynomial
approximation to be more efficient in this second case.

\section{Summary and outlook}

We presented our results for a pilot study of $B_s$-meson inclusive semileptonic decays~\cite{Barone:2023tbl} highlighting the general setup for the calculation. 
In particular, we provide an extended framework that incorporates two different methods, one exploiting Chebyshev polynomials, and one exploiting 
the Backus-Gilbert approach. We showed that the two are compatible for both the computation of the decay rate and the moments,
and outlined how the two regularise the calculation of the observables to reduce the statistical noise, either acting on the data (Chebyshev) or modifying the coefficients
of the polynomial approximation (Backus-Gilbert).

We also tested our setup to address some of the quantities related to hadronic mass and lepton energy moments, similarly to what was done in~\cite{Gambino2022}.
A full treatment of (differential) moments on the lattice will allow to compare with analytical OPE approaches
and will provide a common ground to cross-validate different approaches.
Moreover, moments do not depend on the CKM-matrix elements, and lattice calculations
may allow to extract some of the parameters that appear in the perturbative expansion.

Overall, our work provides a solid foundation for future studies of semileptonic inclusive decays.
However, several aspects require further investigations and will be the subject of future studies,
in particular systematic errors associated with the polynomial approximation~\cite{Kellermann:2022mms}, finite-volume effects, discretisation errors, and the continuum limit.

\section*{Acknowledgments}

This work used the DiRAC Extreme Scaling service at the University of Edinburgh, operated by the Edinburgh Parallel Computing Centre on behalf of the STFC DiRAC HPC Facility (www.dirac.ac.uk). This equipment was funded by BEIS capital funding via STFC capital grant ST/R00238X/1 and STFC DiRAC Operations grant ST/R001006/1. DiRAC is part of the National e-Infrastructure.

The works of S.H. and T.K. are supported in part by JSPS KAKENHI Grant Numbers 22H00138
and 21H01085, respectively, and by the Post-K and Fugaku supercomputer project through the Joint
Institute for Computational Fundamental Science (JICFuS).

%\bibliographystyle{JHEP-notitle}
%\bibliography{biblio}

\providecommand{\href}[2]{#2}\begingroup\raggedright\endgroup

\end{document}